\documentclass[aps,prl,twocolumn,showpacs,superscriptaddress]{revtex4}
\usepackage{graphicx}
\usepackage{dcolumn}
\usepackage{bm}
\usepackage{epsf}%
\usepackage{longtable}%
\usepackage{supertabular}%
\usepackage{float}
\usepackage{threeparttable}
\usepackage[colorlinks = true, citecolor=blue,urlcolor=blue,linkcolor=blue]{hyperref}
\usepackage{color} 

\usepackage{url}
\usepackage{array}
\usepackage{amssymb}

\begin{document}


\title{First direct measurement of an astrophysical $p$ process reaction cross section using a radioactive ion beam}

\author{G. Lotay}
\affiliation{Department of Physics, University of Surrey, Guildford, GU2 7XH, United Kingdom}

\author{S.\ A. Gillespie}
\thanks{Present Address: FRIB, Michigan State University, East Lansing, MI, 48824}
\affiliation{TRIUMF, 4004 Wesbrook Mall, Vancouver, British Columbia, V6T 2A3, Canada}

\author{M. Williams}
\affiliation{TRIUMF, 4004 Wesbrook Mall, Vancouver, British Columbia, V6T 2A3, Canada}
\affiliation{Department of Physics, University of York, Heslington, York, YO10 5DD, United Kingdom}

\author{T. Rauscher}
\thanks{ORCID: 0000-0002-1266-0642}
\affiliation{Department of Physics, University of Basel, Klingelbergstr.\ 82, CH-4056 Basel, Switzerland}
\affiliation{Centre for Astrophysics Research, University of Hertfordshire, Hatfield AL10 9AB, United Kingdom}

\author{M. Alcorta}
\affiliation{TRIUMF, 4004 Wesbrook Mall, Vancouver, British Columbia, V6T 2A3, Canada}

\author{A.\ M. Amthor}
\affiliation{Department of Physics and Astronomy, Bucknell University, Lewisburg, PA, 17837, USA}

\author{C.\ A.\ Andreoiu}
\affiliation{Department of Chemistry, Simon Fraser University, Burnaby, British Columbia V5A 1S6, Canada}

\author{D. Baal}
\affiliation{TRIUMF, 4004 Wesbrook Mall, Vancouver, British Columbia, V6T 2A3, Canada}

\author{G.\ C. Ball}
\affiliation{TRIUMF, 4004 Wesbrook Mall, Vancouver, British Columbia, V6T 2A3, Canada}

\author{S. S. Bhattacharjee}
\altaffiliation{Present Address: Institute of Experimental and Applied Physics, Czech Technical University in Prague, Husova 240/5, 110 00 Prague 1, Czech Republic}
\affiliation{TRIUMF, 4004 Wesbrook Mall, Vancouver, British Columbia, V6T 2A3, Canada}

\author{H. Behnamian}
\affiliation{Department of Physics, University of Guelph, Guelph, Ontario, N1G 2W1, Canada}

\author{V. Bildstein}
\affiliation{Department of Physics, University of Guelph, Guelph, Ontario, N1G 2W1, Canada}

\author{C. Burbadge}
\thanks{Deceased}
\affiliation{Department of Physics, University of Guelph, Guelph, Ontario, N1G 2W1, Canada}

\author{W.\ N. Catford}
\affiliation{Department of Physics, University of Surrey, Guildford, GU2 7XH, United Kingdom}

\author{D.\ T.\ Doherty}
\affiliation{Department of Physics, University of Surrey, Guildford, GU2 7XH, United Kingdom}

\author{N.\ E.\ Esker}
\affiliation{TRIUMF, 4004 Wesbrook Mall, Vancouver, British Columbia, V6T 2A3, Canada}

\author{F.\ H. Garcia}
\affiliation{Department of Chemistry, Simon Fraser University, Burnaby, British Columbia V5A 1S6, Canada}

\author{A.\ B.\ Garnsworthy}
\affiliation{TRIUMF, 4004 Wesbrook Mall, Vancouver, British Columbia, V6T 2A3, Canada}

\author{G. Hackman}
\affiliation{TRIUMF, 4004 Wesbrook Mall, Vancouver, British Columbia, V6T 2A3, Canada}

\author{S. Hallam}
\affiliation{Department of Physics, University of Surrey, Guildford, GU2 7XH, United Kingdom}

\author{K.\ A.\ Hudson}
\affiliation{TRIUMF, 4004 Wesbrook Mall, Vancouver, British Columbia, V6T 2A3, Canada}
\affiliation{Department of Physics, Simon Fraser University, Burnaby, British Columbia V5A 1S6, Canada}

\author{S. Jazrawi}
\affiliation{Department of Physics, University of Surrey, Guildford, GU2 7XH, United Kingdom}

\author{E. Kasanda}
\affiliation{Department of Physics, University of Guelph, Guelph, Ontario, N1G 2W1, Canada}

\author{A.\ R.\ L. Kennington}
\affiliation{Department of Physics, University of Surrey, Guildford, GU2 7XH, United Kingdom}

\author{Y.\ H.\ Kim}
\affiliation{Department of Nuclear Engineering, Hanyang University, Seoul, Republic of Korea}

\author{A. Lennarz}
\affiliation{TRIUMF, 4004 Wesbrook Mall, Vancouver, British Columbia, V6T 2A3, Canada}

\author{R.\ S. Lubna}
\affiliation{TRIUMF, 4004 Wesbrook Mall, Vancouver, British Columbia, V6T 2A3, Canada}

\author{C.\ R.\ Natzke}
\affiliation{TRIUMF, 4004 Wesbrook Mall, Vancouver, British Columbia, V6T 2A3, Canada}
\affiliation{Department of Physics, Colorado School of Mines, Golden, CO 80401, USA}

\author{N. Nishimura}
\affiliation{Astrophysical Big Bang Laboratory, CPR, RIKEN, Wako, Saitama 351-0198, Japan}

\author{B. Olaizola}
\altaffiliation{Present Address: ISOLDE, CERN, CH-1211 Geneva 23, Switzerland}
\affiliation{TRIUMF, 4004 Wesbrook Mall, Vancouver, British Columbia, V6T 2A3, Canada}

\author{C. Paxman}
\affiliation{Department of Physics, University of Surrey, Guildford, GU2 7XH, United Kingdom}
\affiliation{TRIUMF, 4004 Wesbrook Mall, Vancouver, British Columbia, V6T 2A3, Canada}

\author{A. Psaltis}
\altaffiliation{Present Address: Institut f{\"u}r Kernphysik, Technische Universit{\"a}t Darmstadt, Darmstadt,  D-64289, Germany}
\affiliation{Department of Physics and Astronomy, McMaster University, Hamilton, Ontario, L8S 4L8, Canada}

\author{C.\ E.\ Svensson}
\affiliation{Department of Physics, University of Guelph, Guelph, Ontario, N1G 2W1, Canada}

\author{J. Williams}
\affiliation{TRIUMF, 4004 Wesbrook Mall, Vancouver, British Columbia, V6T 2A3, Canada}

\author{B. Wallis}
\affiliation{Department of Physics, University of York, Heslington, York, YO10 5DD, United Kingdom}

\author{D. Yates}
\affiliation{TRIUMF, 4004 Wesbrook Mall, Vancouver, British Columbia, V6T 2A3, Canada}
\affiliation{Department of Physics and Astronomy, University of British Columbia, Vancouver, BC V6T 1Z4, Canada}

\author{D. Walter}
\affiliation{TRIUMF, 4004 Wesbrook Mall, Vancouver, British Columbia, V6T 2A3, Canada}

\author{B. Davids}
\affiliation{TRIUMF, 4004 Wesbrook Mall, Vancouver, British Columbia, V6T 2A3, Canada}
\affiliation{Department of Physics, Simon Fraser University, Burnaby, British Columbia V5A 1S6, Canada}

\date{\today}
\begin {abstract}

We have performed the first direct measurement of the $^{83}$Rb($p,\gamma$) radiative capture reaction cross section in inverse
kinematics using a radioactive beam of $^{83}$Rb at incident energies of 2.4 and $2.7 A$ MeV. The measured cross section at an 
effective relative kinetic energy of $E_{\mathrm{cm}}$ = 2.393 MeV, which lies within the relevant energy window for core collapse 
supernovae, is smaller than the prediction of statistical model calculations. This leads to the 
abundance of $^{84}$Sr produced in the astrophysical $p$ process being higher than previously calculated. Moreover, the discrepancy 
of the present data with theoretical predictions indicates that further experimental investigation of $p$ process reactions 
involving unstable projectiles is clearly warranted.

\end {abstract}

\pacs{25.60.-t, 25.45.Hi, 26.20.Np}

\maketitle

It has long since been established that the stellar nucleosynthesis of elements heavier than iron is largely governed by the
($s$)low and ($r$)apid neutron capture processes \cite{B2FH}. However, there exist $\sim$30 stable, neutron-deficient nuclides, 
between Se and Hg, that cannot be formed by either of the aforementioned processes, and whose astrophysical origin remains a subject 
of active investigation \cite{Rauscher}. These $p$ nuclides, because they account for only a small fraction of overall elemental 
abundances, are not directly observable in stars or supernova remnants. As such, it is necessary to study their formation using a 
combination of detailed nucleosynthetic models and meteoritic data \cite{Rauscher2}.
At present, it is believed that $p$ nuclides are formed by photodisintegration reactions on pre-existing $r$- and $s$-process 
seed nuclei in the O/Ne layers of core-collapse supernovae (ccSN) \cite{Arnould,howard} and in 
thermonuclear supernovae \cite{trav11,nobsnIa}, with typical peak plasma temperatures of $T_{max}\sim2-3.5$ GK in the $p$-process 
layers. In particular, ($\gamma,n$) reactions drive the pathway of nucleosynthesis toward the neutron-deficient side of stability 
until neutron separation energies become high enough that ($\gamma,p$) and ($\gamma,\alpha$) disintegrations largely dominate the 
flow of material. This astrophysical $\gamma$ process is capable of reproducing the bulk of the $p$ nuclides within a single 
stellar site \cite{Rauscher2}. However, there are abiding issues in obtaining abundances consistent with solar system values for 
the lightest $p$ nuclides ($A$ $\lesssim$ 110) \cite{arngor,umberto} to be resolved. In this regard, a possible solution may 
be found in the underlying nuclear physics input, as experimental cross sections of $p$-process reactions are almost entirely 
unknown and the related reaction rates are based entirely on theoretical calculations.  

It is well known (see, e.g., \cite{qval,Rauscher3}) that experimental measurements to constrain stellar rates should be performed 
in the reaction direction of positive $Q$ value, in order to minimize the impact of thermal excitations of target nuclei in the 
stellar plasma and numerical inaccuracies when converting between forward and reverse rates. In the application to the 
nucleosynthesis of $p$ nuclides, this implies that capture reactions instead of the reverse photodisintegration reactions should be 
studied experimentally. The 
vast majority of these reactions involve unstable nuclei and exhibit cross sections of order 100 $\mu$b. As such, most $p$-process 
reactions have remained experimentally inaccessible, even with the latest developments in the production and acceleration of 
radioactive ion beams, and astrophysical abundance calculations have relied extensively on the use of Hauser-Feshbach (HF) theory 
\cite{NONSMOKER1,NONSMOKER2}. Although this approach is valid for reactions appearing in the synthesis of $p$ nuclides, the nuclear 
properties required as input are not well known off stability. This may lead to larger uncertainties in the predictions of 
astrophysical reaction rates and therefore requires experimental validation. 
Consequently, in this Letter, we present the first direct measurement of a $p$-process reaction involving an unstable nuclide, in 
the relevant energy window (Gamow window) for the $\gamma$ process in ccSN ($E_{c.m.}$ $\sim$ 1.4$-$3.3 MeV 
\cite{rauscher3}). 

This pioneering study performed at the ISAC-II facility of TRIUMF utilised an intense radioactive beam of $^{83}$Rb ions, together with 
the TIGRESS $\gamma$-ray array \cite{TIGRESS} and the newly-commissioned EMMA recoil mass spectrometer \cite{EMMA}, to investigate 
the astrophysical $^{83}$Rb($p$,$\gamma$)$^{84}$Sr reaction. In particular, by exploiting the fact that the electromagnetic decay of 
proton-unbound states in $^{84}$Sr, populated via resonant proton capture on the 5/2$^{-}$ ground state of $^{83}$Rb, predominantly 
proceeds via $\gamma$-decay cascades to the lowest-lying 2$^{+}$ level, rather than directly to the ground state, it was possible to 
determine the reaction cross section from the observed 793.22(6)-keV, 2$^{+}_1$ $\rightarrow$ 0$^{+}_1$ $\gamma$-ray yield 
\cite{Singh}. This not only provides valuable information for current models of $p$-process nucleosynthesis but also represents a 
new approach to the direct measurement of astrophysical reaction cross sections. Most notably, the $^{83}$Rb($p$,$\gamma$)$^{84}$Sr 
reaction impacts the $^{84}$Sr abundance obtained in ccSN
\cite{Rapp,Rauscher} and elevated levels of $^{84}$Sr have recently been discovered in calcium-aluminium-rich inclusions (CAIs) in 
the Allende meteorite \cite{Charlier}. Whilst it has been proposed that the $^{84}$Sr abundances found in CAIs may be accounted for 
by $r$- and $s$-process variability in $^{88}$Sr production, such distributions are most easily described by an anomaly in the 
astrophysical $p$ process.

Here, radioactive $^{83}$Rb ions ($t_{1/2}\sim86$ days), produced and accelerated to energies of $2.4$ and $2.7 A$~ MeV by the ISAC-II facility of TRIUMF, were used to bombard 300 to 900 $\mu$g/cm$^2$ thick polyethylene (CH$_2$)$_n$ targets at intensities of $1-5\times10^7$~s$^{-1}$ in order to perform measurements of the $^{83}$Rb($p,\gamma$) reaction cross section. A measurement of the stable $^{84}$Kr($p,\gamma$) radiative capture cross section was carried out as well at a bombarding energy of $2.7A$~MeV in a test of the new experimental setup with a nearly identical mass beam free from radioactive-beam-induced background. The intensities of both the stable and radioactive beams were limited to maintain the integrity of the target foils; much greater intensity on target was available from the ISAC-II accelerator. Prompt $\gamma$ rays were detected with the TIGRESS array, which, in this instance, consisted of 12 Compton-suppressed HPGe detectors \cite{TIGRESS}, while recoiling $^{85}$Rb and $^{84}$Sr nuclei were transmitted to the focal plane of the EMMA recoil mass spectrometer \cite{EMMA} in either the $25^+$ or $26^+$ charge state. The electrodes of the two electrostatic deflectors were held at potential differences of $\sim$320 kV while three slit systems enabled the rate of scattered beam reaching the focal plane to be suppressed by a factor of $\sim$50,000; such beam suppression was required in order to reduce radioactive beam-induced background to a tolerable level but the slit settings did not diminish the transmission efficiency for recoils due to their small angular and energy spreads. An electromagnetic separator capable of a relatively large electrostatic rigidity of 13 MV was needed to transmit the recoils of these reactions. The rigidity limits of EMMA imply that relative kinetic energies up to 10\% larger than studied here can be reached, rendering the spectrometer well matched to the Gamow window for the $p$ process.

Recoils were highly forward focussed due to the inverse kinematics and including the effects of multiple scattering in the target foil emerged at scattering angles not exceeding $0.4^{\circ}$. This allowed for very high recoil transport efficiency which was estimated on the basis of empirical energy and angular acceptance studies with an $\alpha$ source to exceed 99\%. Recoils and scattered beam reaching the focal plane were detected by a parallel grid avalanche counter, a transmission ionization chamber, and a 500 $\mu$m thick ion-implanted Si detector \cite{EMMA}.

The charge state distribution of a reduced-intensity beam of $2.7A$ MeV $^{84}$Kr was measured and used to infer the charge state fractions of $^{85}$Rb and $^{84}$Sr recoils, using the dependence of the equilibrium charge state on $Z$ and energy predicted by the empirical parametrization of Ref.\ \cite{shima}. The intensities of 6 charge states were measured. During the radiative capture cross section measurements the integrated luminosity was obtained by monitoring target protons elastically scattered into two 150 mm$^2$ silicon surface barrier detectors mounted at $20^\circ$ with respect to the beam axis downstream of the target position, relative to regular Faraday cup readings, while $\gamma$-ray detection efficiencies were established using standard $^{152}$Eu and $^{56}$Co sources. During the measurement of the $^{83}$Rb($p,\gamma$) reaction, background arose due to the presence of contaminant $^{83}$Sr in the beam. In particular, $^{83}$Sr scattering into the entrance aperture of EMMA resulted in the detection of $^{83}$Sr $\beta$-delayed $\gamma$ rays in the TIGRESS array. Consequently, immediately following the experiment, and again 22 days later, the GRIFFIN spectrometer \cite{GRIFFIN} was used to study the decay of beam ions scattered into the entrance aperture; the $^{83}$Rb fraction was determined to be 62(3)$\%$. 

Estimates of the relative uncertainties associated with the integrated luminosity, the recoil transmission efficiency, $\gamma$ ray detection efficiency, and charge state fractions amount to $\pm19\%$, $^{+0.1}_{-33}$\%, $\pm5\%$, and $\pm10$\%, respectively. We note that the recoil transmission efficiency is known to be high based on the measured transmission of $^{84}$Kr and $^{83}$Rb beam ions during attenuated beam runs and due to the small recoil cone angle and kinetic energy spread of $\pm1$\%. However, we have placed a very conservative estimate on its lower limit to account for any possible unforeseen losses during the measurement of the ($p,\gamma$) reaction cross sections, given the large energy losses in the thick targets and the unmeasured stopping powers of $^{84}$Sr and $^{85}$Rb ions in polyethylene. The statistical uncertainty in the data acquisition live-time fraction, which exceeded 90\% for data taking with both beams, is negligible.

\begin{figure}[!ht]
\includegraphics[width=\linewidth]{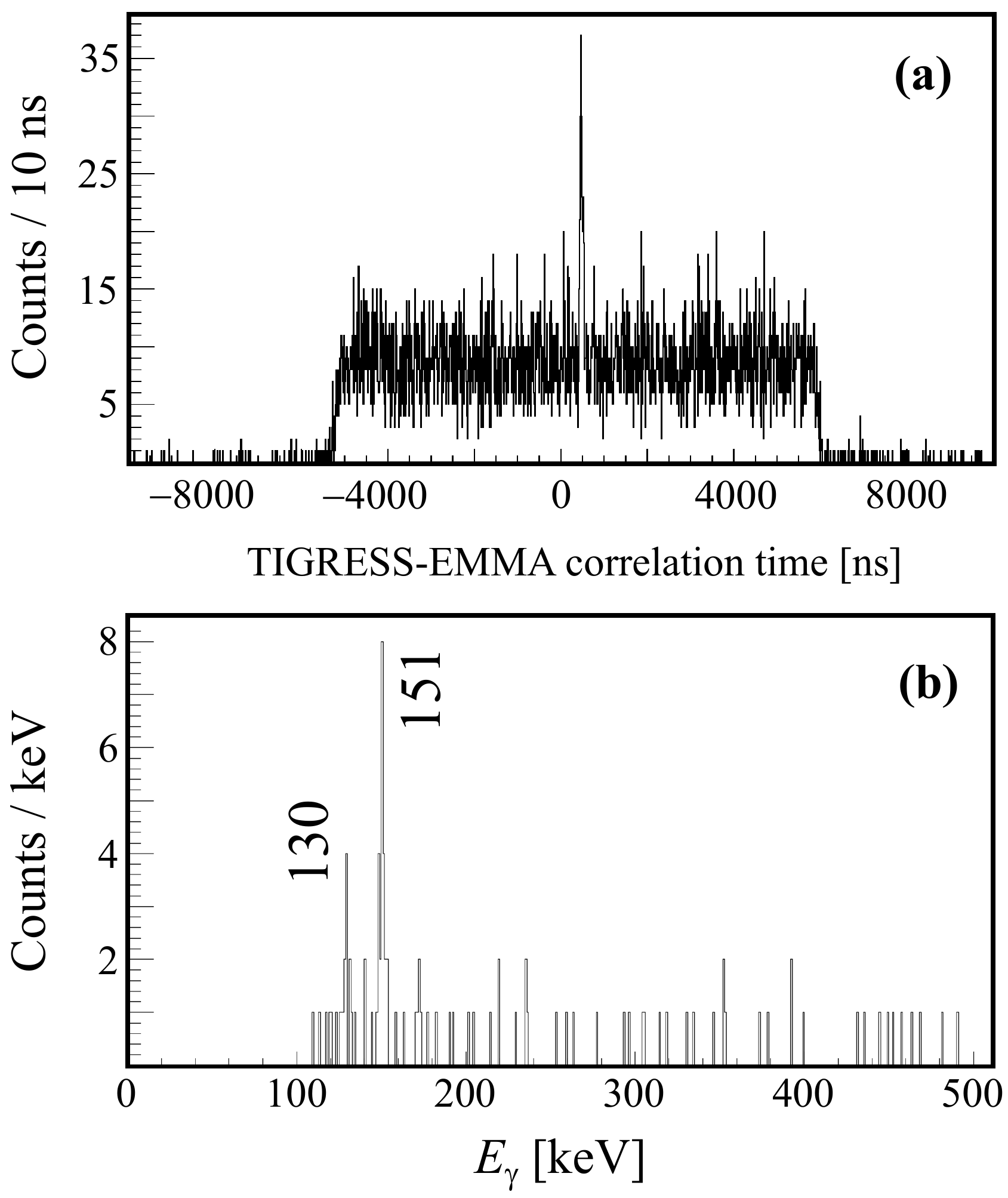}
\caption{\label{fig1}(a) Time difference between events observed in the TIGRESS $\gamma$-ray array and the focal plane of the EMMA recoil mass spectrometer, following the $^{84}$Kr($p,\gamma$) reaction. (b) Energies of $\gamma$ rays observed in coincidence with $A=85$ recoils from the timing peak shown in panel (a).}
\end{figure}

An example of a timing peak observed in this study, corresponding to the time difference between $\gamma$-ray events registered in TIGRESS 
and recoils detected at the focal plane of EMMA, is presented in Fig.\ \ref{fig1}. Such a timing peak provides clear evidence for 
distinct ($p,\gamma$) events and, by placing a software gate on this peak for the measurement of the $^{84}$Kr($p,\gamma$)$^{85}$Rb reaction, 
130- and 151-keV $\gamma$ rays, corresponding to decays from the 1/2$^{-}_1$ and 3/2$^{-}_1$ levels in $^{85}$Rb \cite{Singh2}, were 
cleanly identified (see Fig. 1(b)). In this case, the $1/2^{-}_1$ and $3/2^{-}_1$ excited states were populated following primary 
$\gamma$ decays from high-lying, proton-unbound levels in $^{85}$Rb. As such, the observed $\gamma$-ray intensities provide direct 
measures of the inclusive partial reaction cross sections. Note, e.g., that the $1/2^{-}_1$ state decays 99.42(9)\% of the time to 
the $3/2^{-}_1$ level \cite{Singh2}, so the total radiative capture cross section is not the sum of all the partial cross sections. 
Rather, the total cross section can be inferred from the measured partial cross section and the calculated population of the state 
through a $\gamma$ cascade.

\begin{ruledtabular} 
\begin{table*}[t]
\caption{Parameters used for the determination of radiative capture cross sections. The integrated luminosity represents the product of the total number of beam ions and the areal target density. The detection efficiency is the product of the recoil transmission efficiency, the recoil charge state fraction, the focal plane detection efficiency, the live-time fraction, and the $\gamma$-ray detection efficiency. Upper limits are specified at the 90$\%$ CL. Predicted cross sections are based on a statistical model of the reaction \protect\cite{NONSMOKER2}.}

\begin{tabular}{ccccccccccc}
 
 Reaction &
 $E_{\gamma}$ &
 Transition &
Integrated& 
Events & 
Detection& 
$E_{\mathrm{cm}}$  &
Measured &
Calculated & Measured & Predicted\\
 & & &Luminosity&&Efficiency&&$\sigma_{\mathrm{partial}}$ & Population & $\sigma_{\mathrm{total}}$ & $\sigma_{\mathrm{total}}$\\
  & (keV) &  & ($\mu$$b$$^{-1}$)  & & ($\%$) & (MeV) & ($\mu$$b$) & (\%) & ($\mu$$b$) & ($\mu$$b$) \\ \hline

$^{83}$Rb($p,\gamma$)$^{84}$Sr & 793 & $2^{+} \rightarrow 0^{+}$  & 28(5) & 16(6) & 1.2$^{+0.1}_{-0.4}$ & 2.393 & 49$^{+37}_{-21}$ & 71(10) & 69$^{+54}_{-31}$ & 262 \\ 
 & 793 & $2^{+} \rightarrow 0^{+}$ & 16(2) & $< 16$ & 1.1$^{+0.1}_{-0.4}$ & 2.259 & $< 102$ & 71(10) & $<143$ & 154 \\ 
 
$^{84}$Kr($p,\gamma)^{85}$Rb  & 151 & $3/2^{-} \rightarrow 5/2^{-}$ & 12(2) & 22(5) & 3.1$^{+0.4}_{-1.1}$ & 2.435 & 59$^{+40}_{-18}$ & 65(10) & $91^{+63}_{-31}$ & 385\\ 
& 130 & $1/2^{-} \rightarrow 3/2^{-}$ & 12(2) & 11(4) & 3.1$^{+0.3}_{-1.1}$ & 2.435 & 31$^{+22}_{-12}$ & 27(10) & $115^{+93}_{-62}$ & 385

\end{tabular}
\label{table}

\end{table*}
\end{ruledtabular}

For the measurement of the $^{84}$Kr($p,\gamma$)$^{85}$Rb reaction, an effective relative kinetic energy, $E_\mathrm{cm}^\mathrm{eff}$, of 
2.435 MeV was determined from the incident beam energy ($E_{beam}$ = 2.7$A$ MeV) and energy loss through the (CH$_2$)$_n$ target, 
assuming a reaction cross-section energy-dependence similar to the one obtained from statistical model calculations
\cite{NONSMOKER1,NONSMOKER2}. Specifically, effective energies were calculated by solving Equation 1 for $E_\mathrm{cm}^\mathrm{eff}$.
\begin{equation}\label{eq1}
\langle \sigma(E) \rangle=\frac{\int_{E_f}^{E_i} \sigma(E) dE}{\int_{E_f}^{E_i} dE}=\sigma(E_\mathrm{cm}^\mathrm{eff})
\end{equation}
Target thicknesses were established with an $\alpha$ source and the corresponding energy loss of the beam ($E_i-E_f$) was calculated 
using the programme SRIM \cite{SRIM}. The relative uncertainty in the measured cross section due to the determination of the 
effective energy is estimated to be $\pm16$\%, while the decay branching ratios of 27$\%$ and 65$\%$ to the 1/2$^{-}_1$ and 
3/2$^{-}_1$ excited states in $^{85}$Rb, respectively, are expected to be accurate to within $\pm$10$\%$ (see below). The relative cross section uncertainty due to the 
effective energy determination is estimated via a comparison between the energy dependence of the cross sections predicted by the statistical model at the 
effective energy and at the effective energy calculated with an energy-independent astrophysical $S$ factor. Here, we observe 22(5) 
counts due to the 151-keV $\gamma$-ray transition in $^{85}$Rb, resulting from the  $^{84}$Kr($p,\gamma$) reaction, while 11(4) 
counts are observed from the 130-keV transition that dominates the decay of the 281-keV state. Combining these yields with the 
predicted branching ratios in a weighted average, we infer a total reaction cross section at $E_{\mathrm{cm}}$ = 2.435 MeV of 
$94^{+64}_{-30}$ $\mu$$b$. A summary of the 
parameters used for the determination of the reaction cross sections is given in Table \ref{table}.

\begin{figure}[!ht]
\includegraphics[width=\linewidth]{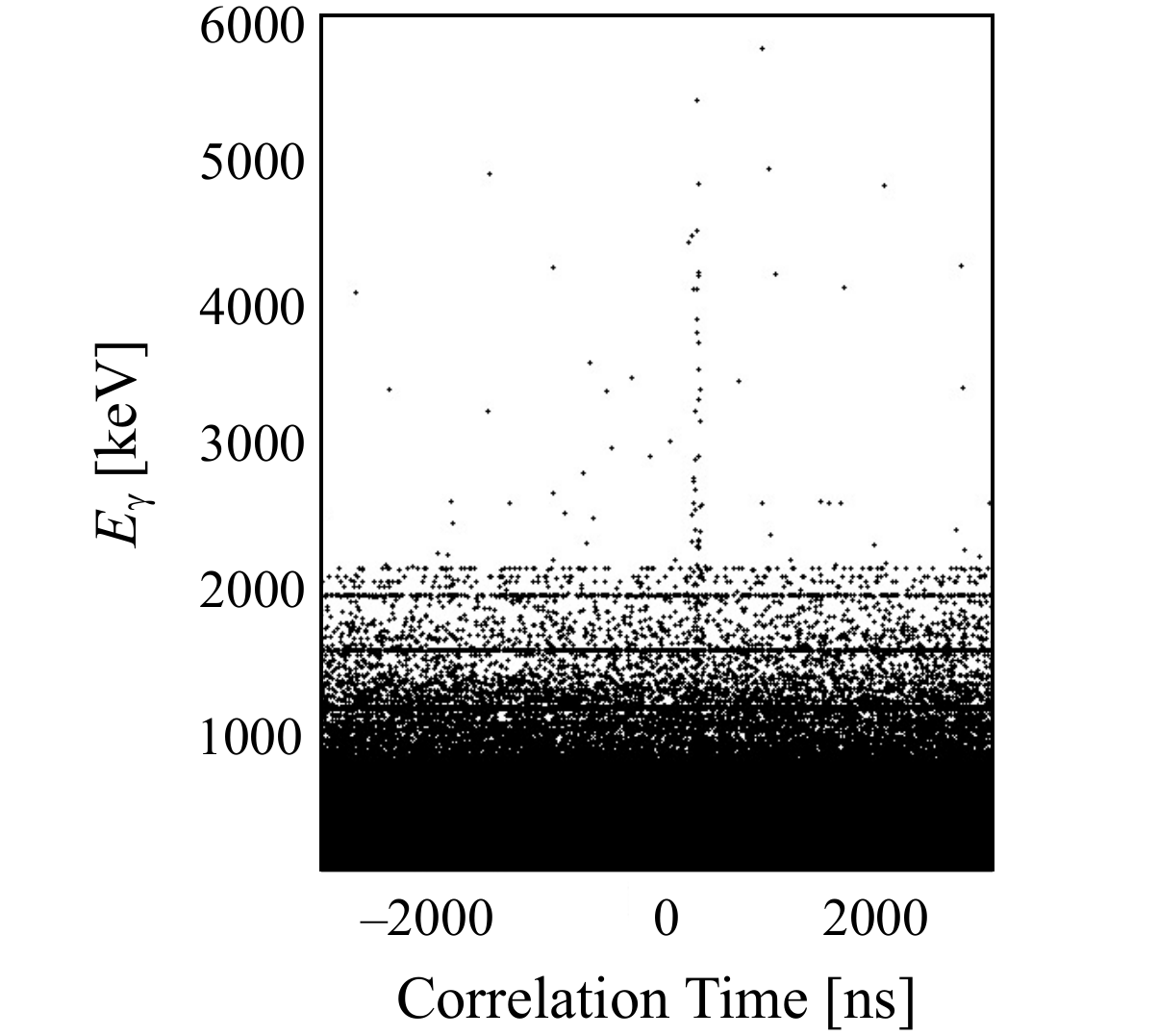}
\caption{\label{fig2}Observed $\gamma$-ray energies in the TIGRESS array, during the measurement of the $^{83}$Rb($p,\gamma$) reaction, as a function of  TIGRESS-EMMA correlation time. A vertical cluster of counts indicates the observation of correlated primary and secondary $\gamma$ rays up to high energies, corresponding to $^{83}$Rb($p,\gamma$) events.}
\end{figure}

In considering the astrophysically important $^{83}$Rb + $p$ reaction, clearly correlated $\gamma$ rays, extending to high energies, are observed at an effective energy of $E_{\mathrm{cm}}$ = 2.393 MeV, as shown in Fig.\ \ref{fig2}. This is entirely consistent with the population of proton-unbound levels in $^{84}$Sr and provides conclusive evidence for $^{83}$Rb($p,\gamma$) events in the present work. However, as can also be seen in Fig.\ \ref{fig2}, there is significant background throughout the low-energy part of the spectrum, due to the $\beta$-delayed $\gamma$ decay of $^{83}$Sr (a known beam contaminant). Nevertheless, it is possible to accurately account for this background using well-known $^{83}$Sr decay data \cite{McCutchan} and by only investigating $\gamma$-decay transitions detected in the 8 detectors centred at 90$^{\circ}$ with respect to the beam axis. In this regard, when applying a Doppler correction appropriate for $^{84}$Sr recoils, $\beta$-delayed transitions from the decays of stopped $^{83}$Sr beam contaminants are shifted into several distinct peaks according to the angles of the detectors, while prompt ($p,\gamma$) transitions are observed as a peak at a single energy. 

\begin{figure}[!ht]
\includegraphics[width=\linewidth]{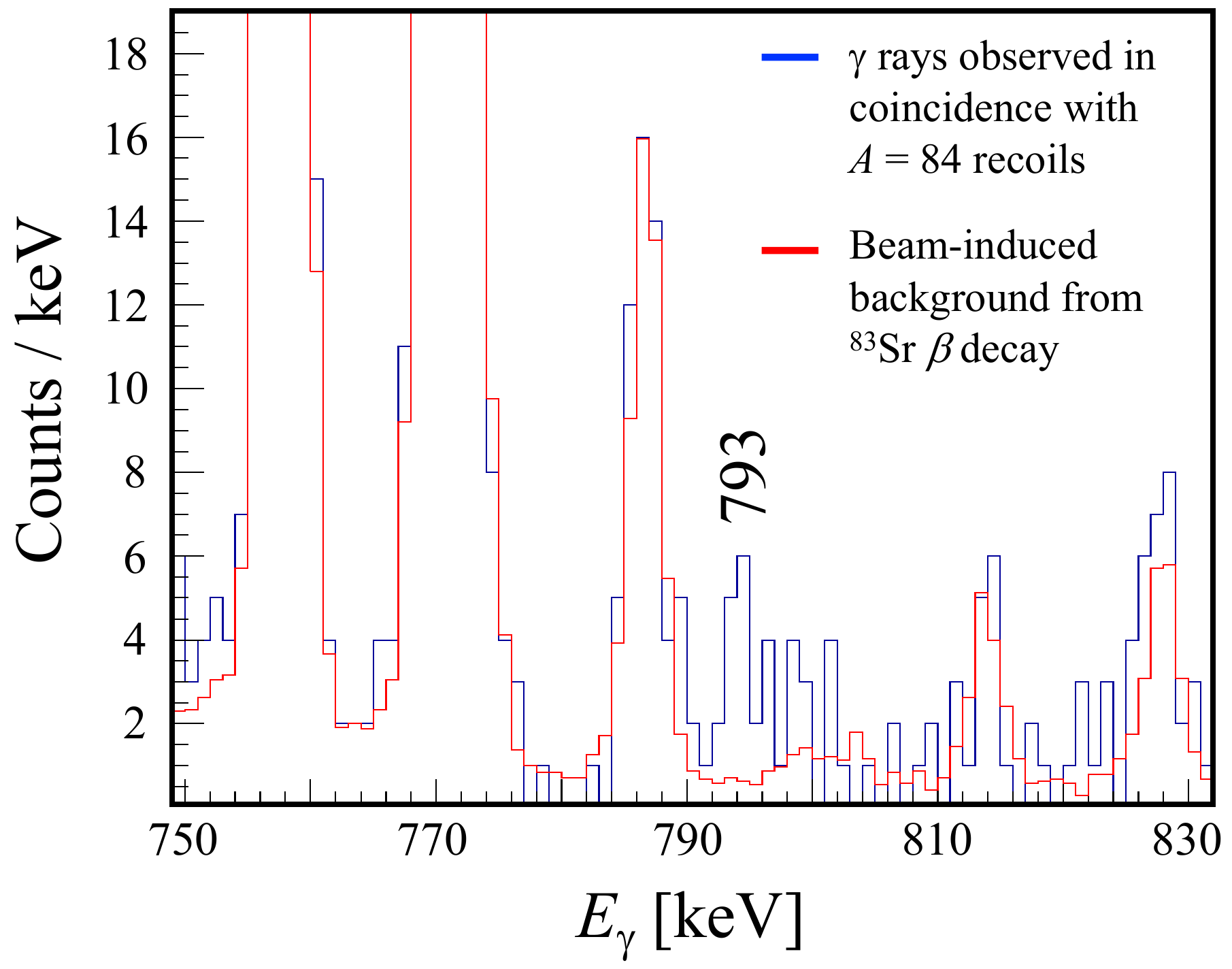}
\caption{\label{fig3}(Color online) Gamma rays observed in the 8 TIGRESS detectors centred at 90$^{\circ}$ with respect to the beam axis in coincidence with $A=84$ recoils, following the $^{83}$Rb($p,\gamma$) reaction.}
\end{figure}

Figure \ref{fig3} illustrates the $\gamma$ decays observed in the 8 TIGRESS detectors centred at 90$^{\circ}$ with respect to the beam axis in coincidence with $A=84$ recoils transmitted to the focal plane of EMMA, during the measurement of the $^{83}$Rb($p,\gamma$) reaction at $E_{\mathrm{cm}}$ = 2.393 MeV. A timing gate 150 ns wide was applied to obtain the coincidence spectrum while the estimated beam-induced background spectrum was obtained using a 1500-ns-wide timing gate on either side of the coincidence peak, correspondingly normalized by a factor of 1/20. Here, 16(6) counts, in excess of those expected as a result of beam-induced background, are observed at 793 keV, indicating strong population of the 2$^{+}_1$ excited level in $^{84}$Sr \cite{Singh}. As such, we measure a partial radiative cross section to the 2$^{+}_1$ excited state in $^{84}$Sr of 49$^{+37}_{-21}$ $\mu$$b$. For inferring the total reaction cross section it is necessary to determine the relative amount of $\gamma$-decays passing through this state. To this end we performed a calculation with the code SMARAGD \cite{SMARAGD,SMARAGDversion}, which is the successor to the NON-SMOKER code \cite{NONSMOKER1,NONSMOKER2} and allows -- additionally to a standard Hauser-Feshbach approach -- to consistently compute level populations through the $\gamma$-cascade in the compound nucleus. Based on this calculation, it is expected that 71(10)$\%$ of the total radiative capture cross section flows through this state and, in the present work, no other decay branches were observed.

A further measurement of the $^{83}$Rb($p,\gamma$) reaction was also performed at $E_{\mathrm{cm}}$ = 2.259 MeV. Regrettably, only a small excess of counts above background was observed at 793 keV in the resultant $\gamma$-ray spectrum, corresponding to population of the 2$^{+}_{1}$ excited state in $^{84}$Sr. Therefore, only an upper limit could be placed on the $^{83}$Rb($p,\gamma$) reaction cross section at $E_{\mathrm{cm}}$ = 2.259 MeV. An upper limit on the signal in the presence of expected background events was derived using the method of Feldman and Cousins \cite{Feldman}, leading to a limit of $< 16$, $\gamma$-gated, $A=84$ recoils at the 90$\%$ confidence level (CL).

\begin{figure}[!ht]
\includegraphics[width=\linewidth]{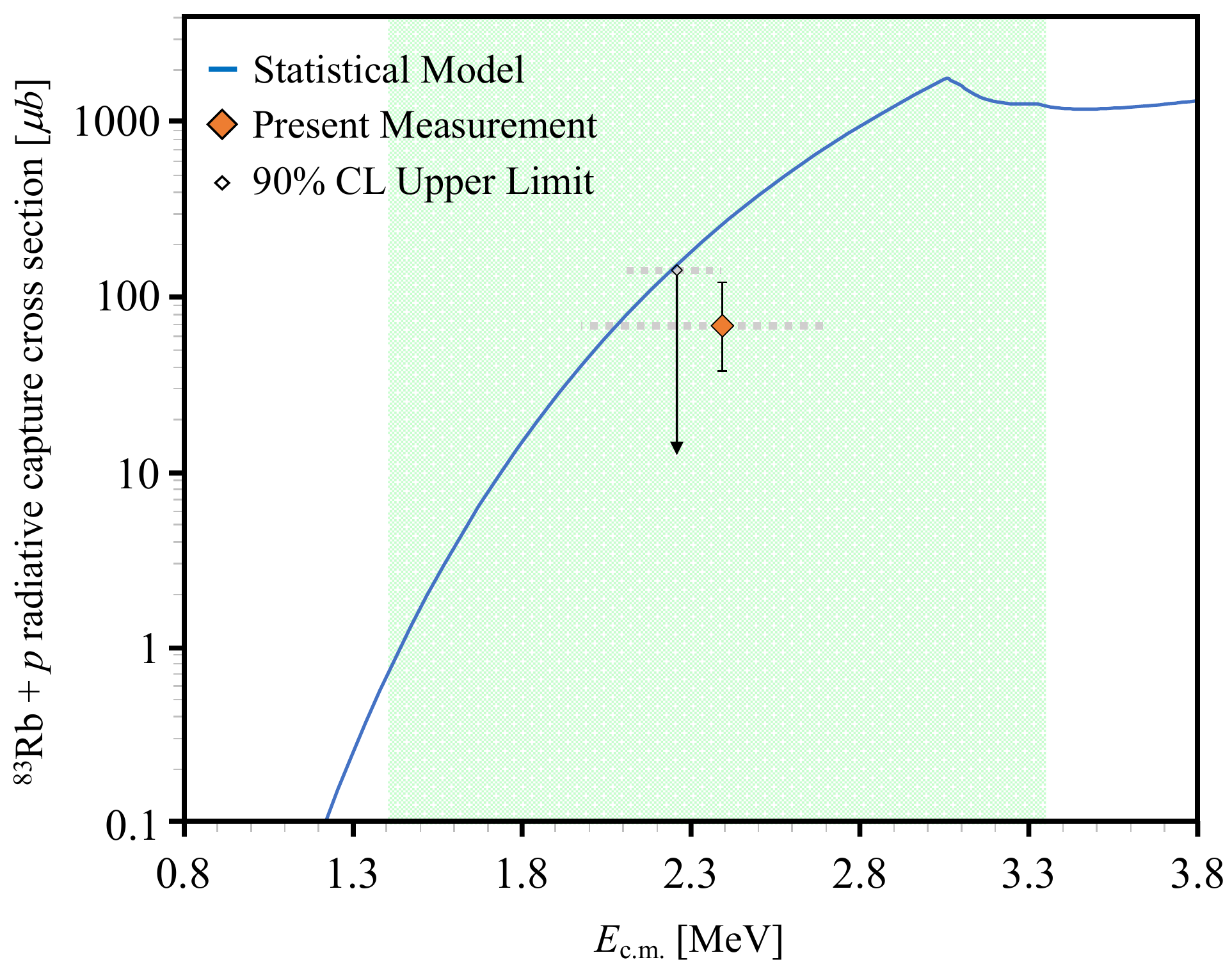}
\caption{(Color online) Total cross section of the $^{83}$Rb($p,\gamma$) reaction inferred from the measured partial cross section 
for populating the 793 keV state in $^{84}$Sr in comparison with statistical model 
predictions \cite{NONSMOKER2}. The shaded region indicates the approximate location of the Gamow window 
\protect\cite{rauscher3} for the $^{83}$Rb($p,\gamma$) reaction in ccSN ($2~\mathrm{GK}<T<3.5~\mathrm{GK}$), the experimental points are centred on the effective relative kinetic energies, and 
the dashed horizontal bars indicate the energies covered in each measurement. The measured point at 
E$_\mathrm{cm}=2.259$~MeV is a 90\% confidence level upper limit.}
\label{fig4}
\end{figure}

In order to assess the astrophysical impact of the present work on $p$-nuclide abundances, in Fig.\ \ref{fig4} we compare the predictions of the statistical model code NON-SMOKER \cite{NONSMOKER1,NONSMOKER2} with the total cross sections inferred from the experimentally measured partial $^{83}$Rb($p,\gamma$) reaction cross sections. The NON-SMOKER results for a wide range of nuclides provide the default set of reaction rates for astrophysical calculations in the absence of experimental data. It is hard to draw strong conclusions from the upper limit at $E_{\mathrm{cm}}$ = 2.259 MeV, but the present experimental 
determination of the cross section of the $^{83}$Rb($p,\gamma$) reaction at $E_{\mathrm{cm}}$ = 2.393 MeV seems to indicate a 
value smaller than the HF prediction and thus implies a reduced thermonuclear reaction rate in comparison to previous 
expectations. A recent study
\cite{Rauscher} reported a strong anti-correlation between the rate of the $^{83}$Rb($p,\gamma$) reaction and the final abundance 
of the $p$ nuclide $^{84}$Sr. In a first, exploratory recalculation using the same approach as in \cite{Rauscher} and assuming a 
reduction of the rate by roughly a factor of four, in line with the cross section range permitted by the experimental results, we 
found an increase in the resulting $^{84}$Sr abundance. This may help explain the observation of enhanced $^{84}$Sr levels in CAIs 
of the Allende meteorite. A more detailed account of the astrophysical simulation will be given in an extended follow-up paper.

In summary, we have performed the first direct measurement of the cross section of an astrophysical $p$ process reaction in the 
Gamow window of ccSN using a radioactive beam. A novel experimental method  allowed us to measure the 
partial cross section of the $^{83}$Rb($p,\gamma$) reaction at energies of $E_{\mathrm{cm}}$ = 2.259 and 2.393 MeV, indicating that 
the thermonuclear reaction rate is
lower than that predicted by statistical model calculations. This is most likely caused by an inaccurately predicted proton width 
\cite{sensi} and requires further investigation using data across a wider energy range. With a smaller reaction cross section, the 
abundance of $^{84}$Sr produced during the astrophysical $p$ process becomes higher than previously expected, offering a 
possible explanation for the observation of elevated levels of $^{84}$Sr discovered in meteorites. Furthermore, given the 
discrepancy between the present experimental measurements and theoretical predictions, we now strongly encourage the further study 
of $p$ process reactions involving unstable projectiles. These reactions may hold the key to understanding the observed abundances 
of several $p$ nuclides throughout our Galaxy.

\begin{acknowledgments}
The authors acknowledge the generous support of the Natural Sciences and Engineering Research Council of Canada. TRIUMF receives 
federal funding via a contribution agreement through the National Research Council of Canada. The GRIFFIN infrastructure was funded 
jointly by the Canada Foundation for Innovation, the Ontario Ministry of Research and Innovation, the British Columbia Knowledge 
Development Fund, TRIUMF, and the University of Guelph. UK personnel were supported by the Science and Technologies Facilities Council (STFC). 
T.R. acknowledges support by the European COST action ``ChETEC" (CA16117). N.N. acknowledges support by JSPS KAKENHI (19H00693, 21H01087).

\end{acknowledgments}


 \normalsize

 \end{document}